\documentclass[11pt,twoside]{article}
\usepackage{asp2010}
\usepackage{bm}

\resetcounters
\begin{document}
\label{authorguide}
\title{The role of quantum interference and partial redistribution in  
the solar Ba~{\sc ii} D$_2$ 4554~\AA\ line}

\author{H. N. Smitha$^{1}$, K. N. Nagendra$^{1}$,
J. O. Stenflo$^{2,3}$ and M. Sampoorna$^{1}$} 
\affil{$^1$Indian Institute of Astrophysics, Koramangala,
Bengaluru, India}
\affil{$^2$Institute of Astronomy, ETH Zurich,
CH-8093 \  Zurich, Switzerland }
\affil{$^3$Istituto Ricerche Solari Locarno, Via Patocchi,
6605 Locarno-Monti, Switzerland}

\begin{abstract}
The Ba~{\sc ii} D$_2$ line at 4554~\AA\ is a good example, where the 
$F$-state interference effects due to the odd isotopes produce
polarization profiles, which are very different from those of the even
isotopes that do not exhibit $F$-state interference. It is therefore
necessary to account for the contributions from the different isotopes
to understand the observed linear polarization profiles of 
this line. In this paper we present radiative transfer modeling with 
partial frequency redistribution (PRD), which is shown to be essential
to model this line. This is because complete frequency redistribution (CRD)
cannot reproduce the observed wing polarization. 
We present the observed and computed 
$Q/I$ profiles at different limb distances. The theoretical profiles 
strongly depend on limb distance ($\mu$) and the model atmosphere which fits the 
limb observations fails at other $\mu$ positions. 
\end{abstract}

\section{Introduction}
\label{smitha-intro}
The interpretation of the Second Solar Spectrum requires taking into account 
a number of physical processes that are not relevant for the intensity 
spectrum.
Some of the lines observed in the Second Solar Spectrum, 
like the Na\,{\sc i} D$_1$ and D$_2$, Cr~{\sc i} triplet at 5206\,\AA,
Ba\,{\sc ii} D$_1$ and D$_2$, Mg~{\sc ii} h and k, and Ca\,{\sc ii} H and K, are affected by
quantum interference between states of different total angular momentum ($J$ or $F$ states).
The importance of quantum interference 
was first demonstrated both observationally and theoretically by 
\citet[][see also \citealt{smitha-1997A&A...324..344S}]{smitha-1980A&A....84...68S}.
An approximate theoretical approach for treating $J$-state interference with partial 
frequency redistribution (PRD) was proposed by 
\cite{smitha-2011ApJ...733....4S,smitha-2013JQSRT.115...46S}.
This theory was then applied to model the linear polarization profiles
of the Cr~{\sc i} triplet at 5204-5208~\AA~ by \cite{smitha-2012A&A...541A..24S}.

\begin{figure}[t]
\begin{center}
\includegraphics[width=10.0cm]{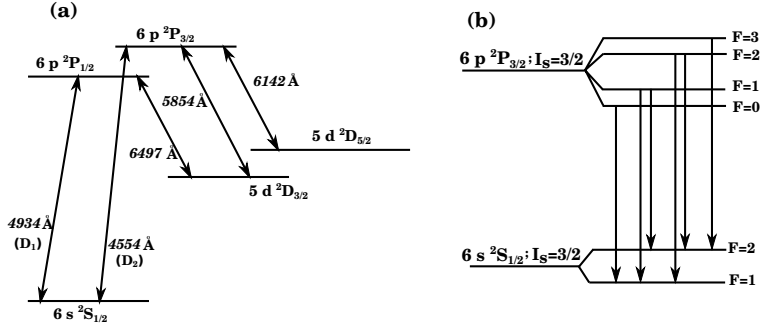}
\caption{(a) the Ba~{\sc ii} 
model atom for the even isotopes. For the odd isotopes
the atomic model is modified by 
replacing two of the levels, $^2P_{3/2}$ and $^2S_{1/2}$, with their hyperfine 
structure components (b). The energy levels are not drawn to the scale.}
\label{smitha-level-diag}
\end{center}
\end{figure}

An atom with a  nuclear spin $I_s$ exhibits hyperfine structure splitting (HFS).
The quantum interference between these hyperfine structure ($F$) 
states produces depolarization in the line core. Na\,{\sc i} D$_2$, 
Ba\,{\sc ii} D$_2$, and Sc\,{\sc ii} line at 4247\,\AA\ are some examples 
of spectral lines whose linear polarization profiles show signatures of $F$-state 
interference. The Ba~{\sc ii} D$_2$ line is due to the 
transition from the upper $J=3/2$ and the lower  
$J=1/2$ fine structure levels (see Figure~\ref{smitha-level-diag}(a)). 
In the odd isotopes of Ba,
both the upper and lower levels undergo HFS due to the nuclear spin $I_s=3/2$. 
This gives rise to four upper and two lower $F$-states
(see Figure~\ref{smitha-level-diag}(b)). In modeling the Ba\,{\sc ii} D$_2$ line
it is necessary to account for quantum interference between the upper $F$-states. 
The odd isotopes constitute about 18\%\ of the total Ba abundance in the Sun
~\citep[c.f. Table 3 of][]{smitha-2009ARA&A..47..481A}.
The rest 82\%\ is from the even isotopes, which are not subject
to HFS (because $I_s=0$). 

High precision spectro-polarimetric observations by 
\cite{smitha-1997A&A...321..927S} using ZIMPOL clearly revealed the existence of three
distinct peaks in the linear polarization $(Q/I)$ profiles of the Ba~{\sc ii} D$_2$ line. 
Using the last scattering approximation,
\cite{smitha-1997A&A...324..344S} explained the physical origin of these 
three peaks. He showed that the central $Q/I$ peak arises from the even isotopes of Ba, 
and the two side peaks from the odd isotopes. 
With a similar last scattering approach, the magnetic
sensitivity of the Ba line was explored by \cite{smitha-2007ApJ...666..588B}.
The above mentioned authors however did not account for radiative transfer
and PRD effects. 
These effects were taken into account by \cite{smitha-2009A&A...493..201F}, but the treatment was limited
to the even isotopes of Ba, for which HFS is absent. In \cite{smitha-2013ApJ...768..163S}
we considered radiative transfer with PRD in 
both even and odd isotopes, and the full effects of HFS with $F$-state interferences. 
Here we present a brief summary of our attempts to model the 
Ba\,{\sc ii} D$_2$ line profile observed in a quiet region close to the 
solar limb. Further we present center-to-limb variation (CLV)
observations of this line and our initial attempts to model the same. 

The paper is organized as follows: In Section~{\ref{smitha-transfer-equations}} we present the 
polarized radiative transfer equation  suitable 
to handle several isotopes of Ba simultaneously. In Section~{\ref{smitha-obs-details}} 
we present the details of the observations. In Section~{\ref{smitha-modeling-pro}}
we discuss the model atom and the model atmosphere used. The results are presented in 
Section~{\ref{smitha-results}} with concluding remarks in Section~{\ref{smitha-conclusions}}. 

\section{Transfer equation for the Ba~{\sc ii} D$_2$ case}
\label{smitha-transfer-equations}
The polarized transfer equation in a  1-D 
non-magnetic axisymmetric atmosphere is given by
\begin{eqnarray}
\mu \frac{{\partial {\bm{\mathcal I}}(\lambda, \mu, z)}
}{\partial z}= 
-k_{\rm tot}(\lambda, z)
\left[{\bm{\mathcal I}}(\lambda, \mu, z) - 
{\bm{\mathcal S}}(\lambda, z)\right],
\label{smitha-transfer}
\end{eqnarray}
in the reduced Stokes vector basis {\citep[see][]{smitha-2012A&A...541A..24S}}.
Positive $Q$ represents the linear polarization oriented parallel to the solar limb.
The total opacity
$k_{\rm tot}(\lambda, z)= k_{l}(z)\phi_g(\lambda,z) + 
\sigma_c(\lambda, z) + k_{\rm th}(\lambda, z)$, with  $\sigma_c$ and 
$k_{\rm th}$ being the continuum scattering and continuum absorption
coefficients, respectively and $k_{l}$ being the wavelength averaged absorption 
coefficient for the $J_a \to J_b$ transition.
Here $J_a$ and $J_b$ are the lower and upper level electronic angular momentum quantum numbers 
respectively. For the case of Ba~{\sc ii} D$_2$, 
the appropriate expression for the profile function $\phi_g$  is 
 \begin{equation}
\phi_g(\lambda,z) = 0.822 ~\phi_e(\lambda,z) + 0.178 ~\phi_o(\lambda,z).
\label{smitha-phi-g}
\end{equation}
Here the Voigt profile function for the even isotopes of Ba~{\sc ii} is denoted  
as $\phi_e(\lambda,z)$. This corresponds to the $J_a=1/2 \to J_b=3/2$ transition 
in the absence of HFS. $\phi_o(\lambda,z)$ is
the profile function for the odd isotopes. 
It is the weighted sum of individual 
Voigt profiles $\phi(\lambda_{F_bF_a},z)$ representing 
each of the $F_a \to F_b$ absorption transition. The initial and the intermediate 
hyperfine split levels have total angular momentum quantum numbers $F_a$ and $F_b$ 
respectively. 
$\phi_o(\lambda,z)$  
is given by
\begin{eqnarray}
\phi_o(\lambda,z)&=& \bigg[\frac{2}{32}\phi(\lambda_{0\,1},z)+ 
\frac{5}{32}\phi(\lambda_{1\,1},z)+\frac{5}{32}\phi(\lambda_{2\,1},z) \nonumber \\ &&+ 
\frac{1}{32}\phi(\lambda_{1\,2},z)+ \frac{5}{32}\phi(\lambda_{2\,2},z)+
\frac{14}{32}\phi(\lambda_{3\,2},z)\bigg]. 
\label{smitha-prof-combined}
\end{eqnarray}
In Equation~(\ref{smitha-phi-g}) the  
contributions from both the $^{135}$Ba (6.6\%) and $^{137}$Ba (11.2\%) odd isotopes
add upto the $17.8\%$ of $\phi_o(\lambda,z)$.
In Equation~(\ref{smitha-transfer}), $\bm{\mathcal S}(\lambda, z)$ is the 
reduced total source vector for a two-level atom with an unpolarized lower level
and is defined as
\begin{eqnarray}
\bm{\mathcal S}(\lambda, z)=&&\frac{k_l(z)\bm{\mathcal S}_{l}(\lambda, z)
+\sigma_c(\lambda, z) \bm{\mathcal S}_{c}(\lambda, z)+
k_{\rm th}(\lambda, z) \bm{\mathcal S}_{\rm th}(\lambda, z)}
{k_{\rm tot}(\lambda, z)}\nonumber \\ && + \ 
\frac{\epsilon~k_l(z)\phi_g(\lambda, z)\bm{\mathcal S}_{\rm th}(\lambda, z)}
{k_{\rm tot}(\lambda, z)}.
\label{smitha-s-total}
\end{eqnarray}
\cite{smitha-2008A&A...481..845D} showed that, for the case of Ba~{\sc ii} D$_2$ 
any ground level polarization would be destroyed by elastic collisions with 
hydrogen atoms \citep[see also][]{smitha-2009A&A...493..201F}.

In Equation~(\ref{smitha-s-total}), $\bm{\mathcal S}_{\rm th}=(B_{\lambda},0)^T$, 
where $B_{\lambda}$ is
the Planck function. The line source vector ${\bm{\mathcal  S}}_{l}(\lambda, z)$ is
\begin{eqnarray}
{\bm{\mathcal  S}}_{l}(\lambda, z)&=&
\int_{0}^{+\infty}\frac{1}{2}
\int_{-1}^{+1} 
{\widetilde{{\bm {\mathcal R}}}(\lambda ,\lambda^{\prime}, z) }
\hat\Psi(\mu'){\bm{\mathcal I}}
(\lambda', \mu', z)\,d\mu'\,d\lambda',
\label{smitha-irreducible-sl}
\end{eqnarray}
with
\begin{equation}
 {\widetilde{{\bm {\mathcal R}}}(\lambda ,\lambda^{\prime}, z)}=
 0.822~\widetilde{{\bm {\mathcal R}}_{e}}(\lambda ,\lambda^{\prime}, z)
+ 0.178~\widetilde{{\bm {\mathcal R}}_{o}}(\lambda ,\lambda^{\prime}, z).
\label{smitha-comb-rm}
\end{equation}
Here $\widetilde{{\bm {\mathcal R}}_{o}}(\lambda,\lambda^{\prime}, z)$ 
is a $(2\times 2)$ diagonal matrix, including the effects of HFS
in odd isotopes.  $\widetilde{{\bm {\mathcal R}}_o}=$ 
diag $({\mathcal R}_o^{0},{\mathcal R}_o^{2})$, 
with ${\mathcal R}_o^{K}$ being the redistribution 
function components for the multipolar index $K$, containing 
both type-II and type-III redistribution of \cite{smitha-1962MNRAS.125...21H}. 
The quantum number replacement  $L\rightarrow J;\ \ \ J\rightarrow F;\ \ \ S\rightarrow I_s$
in Equation~(7) of \citet[][see also \citealt{smitha-2012ApJ...758..112S} and 
\citealt{smitha-2013JQSRT.115...46S}]{smitha-2012A&A...541A..24S}
gives the expression for ${\mathcal R}^{K}$.
We use CRD functions in place of 
the type-III redistribution functions as in \cite{smitha-2013ApJ...768..163S}. 
The contributions from the  
individual redistribution matrices for the odd isotopes $^{135}$Ba and $^{137}$Ba 
are also taken into account.

$\widetilde{{\bm {\mathcal R}}_{e}}(\lambda,\lambda^{\prime}, z)$ 
is also a $(2\times 2)$ diagonal matrix for the even isotopes without HFS.
Its elements ${\mathcal R}_e^{K}$ are the redistribution functions
corresponding to the scattering transition $J_a=1/2 \to J_b=3/2 \to J_f=1/2$.
They are obtained by setting the nuclear spin
$I_s=0$ in  $\widetilde{{\bm {\mathcal R}}_o}(\lambda,\lambda',z)$.
An expression for $\widetilde{{\bm {\mathcal R}}_{e}}(\lambda,\lambda^{\prime}, z)$
in the Stokes vector basis can be found in \cite{smitha-1988ApJ...334..527D} and in 
\citet[][see also \citealt{smitha-1994ApJ...432..274N},
~\citealt{smitha-2011ApJ...731..114S}]{smitha-1997A&A...328..706B}. 

We here use the angle-averaged versions of these quantities.
\cite{smitha-2013MNRAS.429..275S} showed that in the non-magnetic case
the use of the angle-averaged redistribution matrix is sufficiently
accurate for all practical purposes. 
 
We use the two branching ratios \citep[see][]{smitha-2012ApJ...758..112S,smitha-2013ApJ...768..163S}
\begin{eqnarray}
A=\frac{\Gamma_{R}}{\Gamma_{R}+\Gamma_{I}+\Gamma_{E}},\quad
B^{(K)}=\frac{\Gamma_{R}}{\Gamma_{R}+\Gamma_{I}+D^{(K)}}\frac{\Gamma_{E}-D^{(
K)}}{\Gamma_{R}+\Gamma_{I}+\Gamma_{E}},
\label{smitha-a-bk}
\end{eqnarray}
where $\Gamma_R$ and $\Gamma_I$ are 
the radiative and inelastic collisional rates, respectively.
$\Gamma_E$ is the elastic collision rate computed from 
 \cite{smitha-1998MNRAS.300..863B}.
The depolarizing 
elastic collision rates are given by $D^{(K)}$ with $D^{(0)}=0$. 
The $D^{(2)}$ is computed
using \citep[see][]{smitha-2008A&A...481..845D,smitha-2009A&A...493..201F}
\begin{eqnarray}
D^{(2)}&=&6.82\times 10^{-9}n_{\rm H}(T/5000)^{0.40} \nonumber \\ &&
+\  7.44\times10^{-9}(1/2)^{1.5}n_{\rm H}(T/5000)^{0.38}\exp(\Delta E/kT),
\end{eqnarray}
where $n_{\rm H}$ is the neutral hydrogen number density in (cm${^{-3}}$), $T$ the temperature in kelvin,
and $\Delta E$ the energy difference between the $^2P_{1/2}$ and $^2P_{3/2}$
fine structure levels in (cm$^{-1}$) and $k$ is the Boltzmann constant in (cm$^{-1}$/K).
We neglect the collisional coupling between the 
$^2P_{3/2}$ level and the metastable $^2D_{5/2}$ level \citep[see][]{smitha-2013ApJ...768..163S}.
\cite{smitha-2008A&A...481..845D} pointed out the importance of $^2P_{3/2}$ - $^2D_{5/2}$  collisions 
for the line center polarization of Ba~{\sc ii} D$_2$. He showed that the neglect of such collisions would lead to an 
overestimate of the line core polarization by $\sim 25\%$. 
It was shown by \cite{smitha-2009A&A...493..201F} that this in turn
would lead to an overestimate of the microturbulent magnetic field $(B_{\rm turb})$ by $\sim 35\%$. 
Our aim is to explore the roles of PRD, HFS, quantum interferences, and the atmospheric
temperature structure in the modeling of the triple peak structure of the Ba~{\sc ii} D$_2$ linear polarization 
profile.

We assume frequency coherent scattering 
in the continuum \citep[see][]{smitha-2012A&A...541A..24S} with its source vector given by
\begin{equation}
{\bm{\mathcal  S}}_{c}(\lambda, z)= \frac{1}{2}\int_{-1}^{+1} \hat\Psi(\mu')
{\bm{\mathcal I}}(\lambda, \mu', z)\,d\mu'.
\label{smitha-irreducible-sc}
\end{equation}
The matrix $\hat\Psi$ is the Rayleigh scattering phase matrix in the 
reduced basis \citep[see][]{smitha-2007A&A...476..665F}.
$\epsilon$ is the line thermalization parameter defined by
$\epsilon={\Gamma_I}/(\Gamma_R+\Gamma_I)$. 
The Stokes vector $(I,Q)^{\rm T}$ can be computed from the 
irreducible Stokes vector $\bm{\mathcal I}$ by simple transformations
\citep[see][]{smitha-2007A&A...476..665F}.
\section{Observations}
\label{smitha-obs-details}
\begin{figure}[t]
\begin{center}
\includegraphics[width=5.6cm]{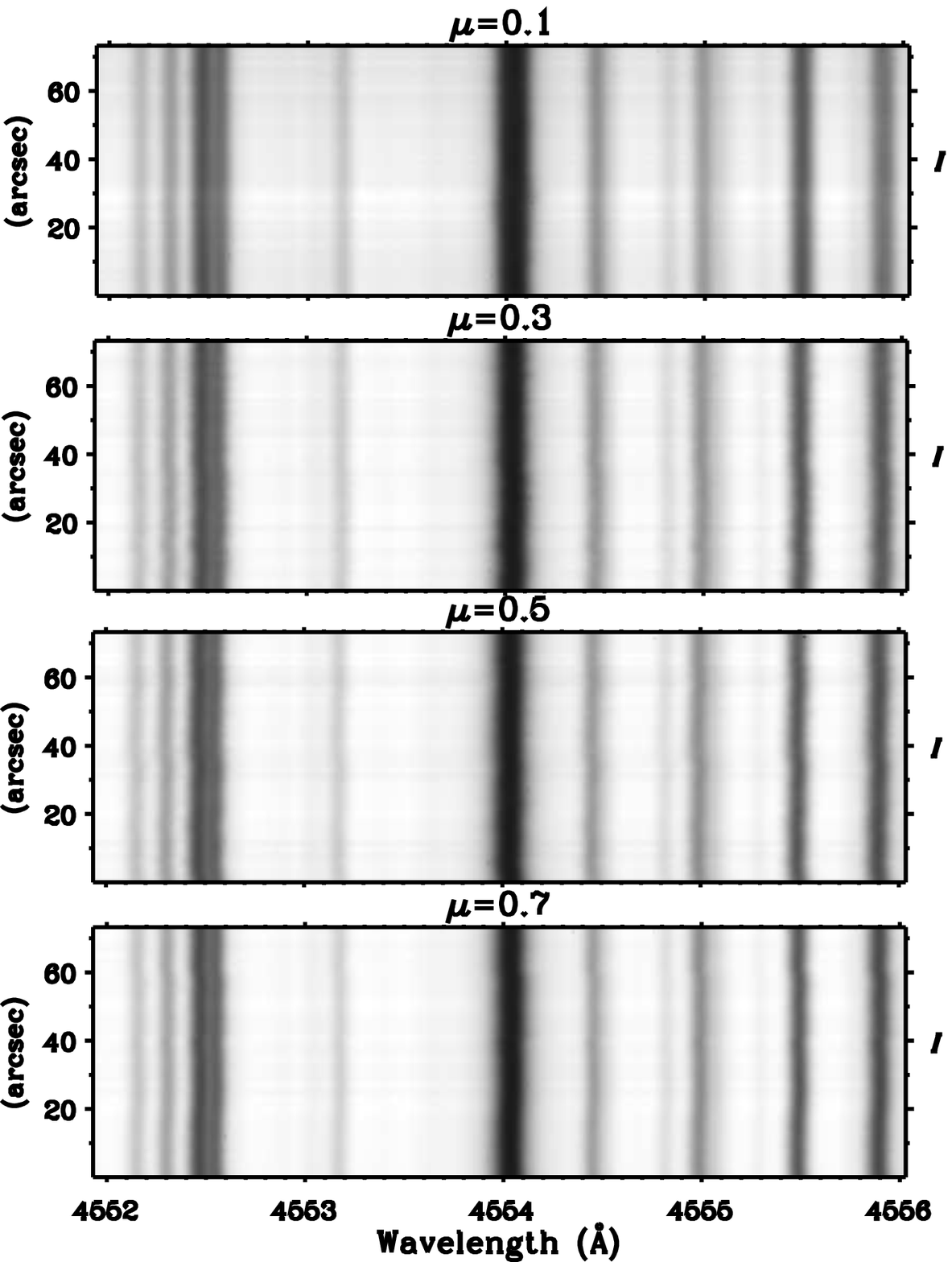}
\ \ \includegraphics[width=5.8cm]{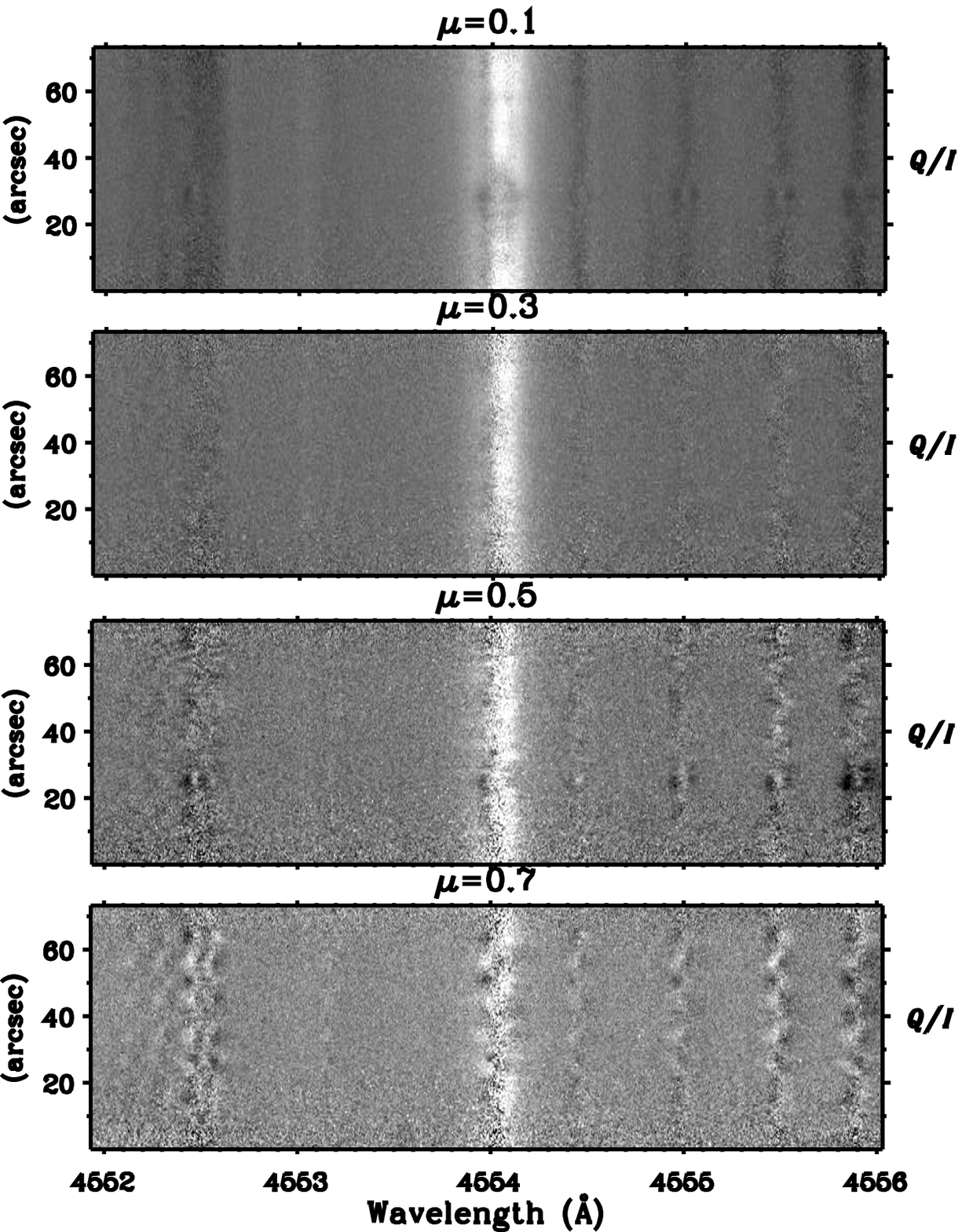}
\caption{CCD images of $I$ and $Q/I$ observed at different limb distances on the solar disk.}
\label{smitha-ccd-images}
\end{center}
\end{figure}

The observed polarization profiles of the Ba~{\sc ii} D$_2$ line shown in this paper 
for $\mu=0.1$ are same as those presented in \cite{smitha-2013ApJ...768..163S}. 
The observations for all $\mu$ values were 
acquired by the ETH team of Stenflo on 
June 3, 2008,  using their ZIMPOL-2 imaging polarimetry system
\citep{smitha-2004A&A...422..703G} at the THEMIS telescope on Tenerife.  
The polarization modulation was done
using Ferroelectric Liquid Crystal modulators.
The observations were recorded at various limb distances
with the slit being placed parallel to the heliographic north pole.
Figure~{\ref{smitha-ccd-images}} shows the CCD images of $I$ and $Q/I$ for different $\mu$ values.
These essentially represent the CLV of the intensity and degree of linear 
polarization. In the Section~\ref{smitha-results} we attempt to model these CLV 
observations. 

\section{Model atom and model atmosphere}
\label{smitha-modeling-pro}
The observed $(I,Q/I)$ profiles of the Ba~{\sc ii} D$_2$ line can be modeled using
a two-stage process. This process is similar to the
one described in \citet[][see also \citealt{smitha-2010ApJ...718..988A,smitha-2011ApJ...737...95A},
~\citealt{smitha-2012A&A...541A..24S,smitha-2013ApJ...768..163S}]{smitha-2005A&A...434..713H}. 
In this two-stage process, the intensity is first computed 
which is then used as an input to compute the polarization.
The code that is used for computing the intensity was developed by 
\citet[][referred to as the RH-code]{smitha-2001ApJ...557..389U}.
To compute the intensity and polarization we need a model atom 
that is representative of the atom under consideration and a one-dimensional 
model atmosphere that mimics the solar atmosphere.
For the case of Ba~{\sc ii} D$_2$, since the effects of different isotopes are to be taken into account, 
we construct three different atom models. Out of these, two are for the odd 
and one for the even isotope. 
The atom model for the even isotope ($^{138}$Ba) is given by the 
five levels of Figure~{\ref{smitha-level-diag}(a)},
while for the odd isotopes ($^{135}$Ba and $^{137}$Ba) the model 
is extended to include the hyperfine splitting as
described by Figure~{\ref{smitha-level-diag}(b)}. The HFS of the 
 $^2P_{1/2}$ level and the metastable levels are neglected. The effects of HFS
are taken into account only for the D$_2$ transition in the intensity computations. 
For the odd isotopes, the radiative transitions between $^2P_{3/2}$ and the metastable levels 
are treated in the same way as for the even isotope. However, in the polarization 
computations, under the two-level atom approximation, the coupling 
to the metastable levels are neglected for both even and odd isotopes.
We neglect the contribution from other less abundant even isotopes.
The wavelengths of the six hyperfine transitions
for the odd isotopes are taken from Kurucz' database and are also
listed in Table~1 of \cite{smitha-2013ApJ...768..163S}.
The line strengths of these transitions are given in Equation~(\ref{smitha-prof-combined}).

\begin{figure}[t]
\begin{center}
\includegraphics[width=10.0cm]{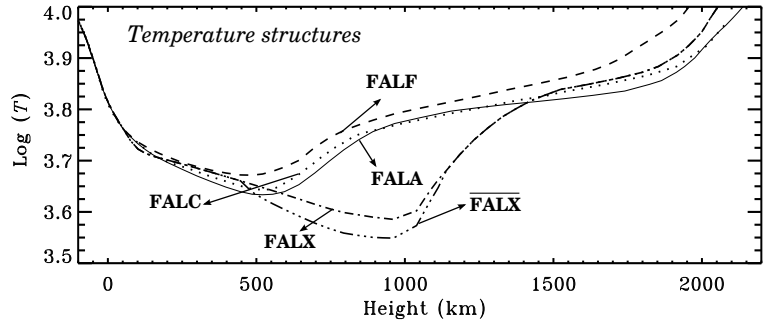}
\caption{The temperature structures of the standard FALA, FALC, FALF, FALX, and 
the modified $\overline{\rm FALX}$ model atmospheres.}
\label{smitha-temp-struc}
\end{center}
\end{figure}

Some of the commonly used standard 1-D solar model atmospheres are the 
FALA, FALF, FALC \citep[see][]{smitha-1993ApJ...406..319F} and the FALX \citep[see][]{smitha-1995itsa.conf..303A}.
Among these four models FALF is the hottest and FALX the coolest. 
However, as will be discussed below, we find that a model atmosphere
that is cooler than FALX is needed to fit the observed profiles. 
The new model, denoted $\overline{\rm FALX}$, is obtained 
by reducing the temperature of the FALX model by about 300 K in the
height range 500 -- 1200 km above the photosphere. The temperature
structures of the standard model atmospheres and the modified model 
are shown in Figure~{\ref{smitha-temp-struc}}.

\section{Results}
\label{smitha-results}
\subsection{Modeling the limb observations}
The three atom models of Ba~{\sc ii} atom described in the above section
give us three different sets of physical quantities
(two for the odd isotopes and one for the even isotope) 
when used in the RH code to compute intensity. 
These quantities are the line opacity, line emissivity, 
continuum absorption coefficient, continuum emissivity, continuum
scattering coefficient, and the mean intensity. 
See \cite{smitha-2001ApJ...557..389U} for the mathematical expressions 
that are used in the RH-code to compute these quantities for the even isotopes.
For the odd isotopes, the profile functions in these expressions
are replaced by $\phi_o(\lambda,z)$ defined by Equation~(\ref{smitha-prof-combined}).

\begin{figure}[t]
\begin{center}
\includegraphics[width=7.0cm]{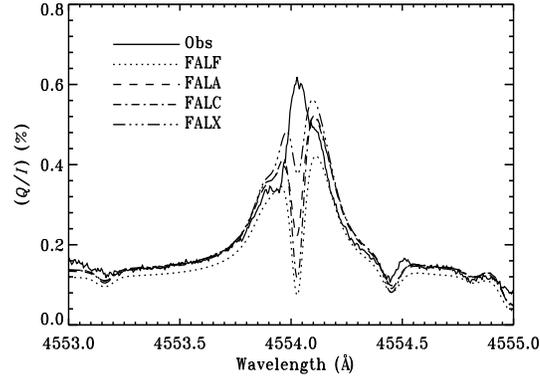}
\caption{Comparison between the observed and theoretical $Q/I$ profiles computed 
using different model atmospheres.}
\label{smitha-no-fit}
\end{center}
\end{figure}

The three sets of quantities obtained from three atom models are combined in the 
ratio of their respective isotope abundances.
The combined quantities are then used as inputs to the polarization code.
The $Q/I$ profiles so computed for the various model atmospheres are shown in Figure~\ref{smitha-no-fit}. 
The figure shows that the $Q/I$ profiles are quite sensitive to the temperature structure
of the model atmosphere and that the central peak cannot be reproduced
from any of the standard model atmospheres. All the models produce a dip  in $Q/I$
at the line center. Earlier studies have shown that such a central dip in $Q/I$
is generally due to the effects of PRD (see for example, 
Figure~3 of \citealt{smitha-1992A&A...258..521F}, and Figure~6 of \citealt{smitha-2005A&A...434..713H}). 
The main contribution to this dip comes from the type-II frequency
redistribution.
 In the case of Ba~{\sc ii} D$_2$, the central  dip is formed mainly through 
contributions from the even isotopes \citep[see Figure~5 of][]{smitha-2013ApJ...768..163S}.
However the profiles computed using only CRD do not show any central dip. 
It is generally known that $Q/I$ profiles computed with CRD and PRD 
in realistic model atmospheres differ not only in the line wings but also in 
the line core \citep[see for e.g.,][]{smitha-1992A&A...258..521F,smitha-2005A&A...434..713H, 
smitha-2009A&A...493..201F,smitha-2012A&A...541A..24S}. This is contrary 
to the behaviour expected of $Q/I$ profiles computed in isothermal atmospheres, 
where they nearly match. A study of the mechanisms that cause the differences 
in the line core, in the case of realistic atmospheres, would be taken up 
in a forthcoming publication.

\cite{smitha-2005A&A...434..713H} show that the magnitude of the $Q/I$ 
central dip is strongly dependent on the choice of atmospheric parameters. 
The central dip becomes shallower for cooler models. 
This behavior is also evident from our Figure~\ref{smitha-no-fit}. 
Such dips in $Q/I$ are common in line profile modeling with PRD, and are often 
smoothed out by applying instrumental and macro-turbulent
broadening. The profiles presented in Figure~\ref{smitha-no-fit} have
already been smeared with a Gaussian 
function having a full width at half maximum (FWHM) of 70~m\AA. Though it is possible 
to smoothen out the dip by increasing the smearwidth, it will suppress 
the side peaks which are due to the odd isotopes and
also disturbs the fit to the intensity profile. 
A smearwidth value of 70~m\AA\ has been chosen by us 
to optimize the fit and it is also consistent with what we expect 
based on the observing parameters and turbulence in the chromosphere.

\begin{figure}[t]
\begin{center}
\includegraphics[width=10.0cm]{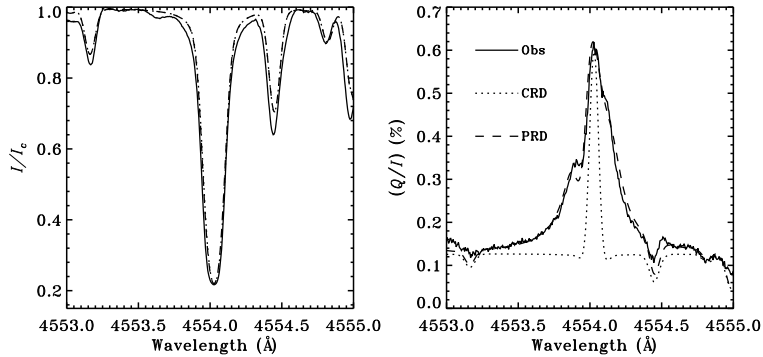}
\caption{Fit to the observed $Q/I$ profiles using the 
$\overline{\rm FALX}$ model atmosphere and also the comparison between 
profiles computed with PRD and CRD.}
\label{smitha-fit}
\end{center}
\end{figure}

Since all the standard model atmospheres failed to provide a fit,
we had to modify the temperature structure of the FALX model to get a new much cooler
model. This modified model is called the  $\overline{\rm FALX}$ 
and the details of its construction are given in \cite{smitha-2013ApJ...768..163S}.
This new model atmosphere succeeds in fitting
both the intensity and the linear polarization profiles. The fit to the $Q/I$ profile is shown as the dashed line 
in Figure~\ref{smitha-fit}.
The theoretical profile in this figure also
includes the contribution from the spectrograph stray light which is 4\% of the continuum intensity.
This is done to mimic the observed profiles which contain the contribution from stray light. 
Finally, the effect of microturbulent magnetic field with a strength of 2~G is included 
to obtain a good $Q/I$ fit at the line center. 

The Ba~{\sc ii} D$_2$ 4554~\AA\ line is a resonance line 
and is strongly influenced by the effects of PRD. 
Hence it is very important to take account of the PRD effects while modeling.
Its importance in case of the even isotopes of Ba~{\sc ii} is demonstrated in \cite{smitha-2009A&A...493..201F}.
In the right panel of Figure~{\ref{smitha-fit}}, we present a comparison between the observed and the theoretical 
$Q/I$ profiles based on CRD alone (dotted line) and on full PRD (dashed line) for
the $\overline{\rm {FALX}}$ model. The intensity profile can be fitted well using either PRD or
CRD \citep[see][]{smitha-2013ApJ...768..163S}, but the polarization profile cannot be fitted 
with CRD alone. This shows that the effects of PRD are essential to model 
the $Q/I$ profiles of the Ba~{\sc ii} D$_2$ line.

\subsection{Modeling the CLV observations}
\begin{figure}[t]
\begin{center}
\includegraphics[width=10.0cm]{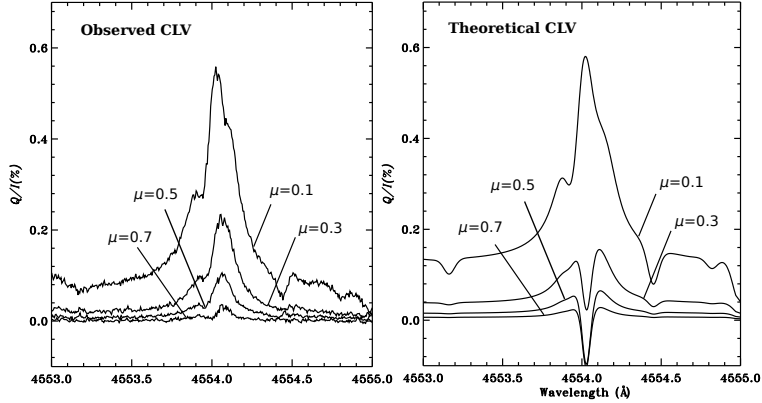}
\caption{Comparison between the observed center to limb variation (CLV) of the 
$Q/I$ profiles and the theoretically computed CLV. The theoretical profiles are
computed using the $\overline{\rm FALX}$ model atmosphere.}
\label{smitha-clv}
\end{center}
\end{figure}

The left panel of Figure~\ref{smitha-clv} shows the CLV of the observed $Q/I$ 
profiles. The right panel of this figure shows the CLV of the theoretical $Q/I$ 
profiles computed
using the $\overline{\rm FALX}$ model atmosphere. As seen from the right panel, the 
newly constructed model atmosphere, though successfully fits the observed profiles for 
$\mu=0.1$, fails to fit the observations for $\mu>0.1$. In fact 
for $\mu>0.1$, we again obtain a central dip instead of a peak in the $Q/I$ profiles.
The central dip
becomes deeper as $\mu \rightarrow 1$. This behavior was noticed also for the case 
of Na~{\sc i} D$_2$ by \cite{smitha-2005A&A...434..713H}. Thus the modification in the temperature 
structure of the FALX model atmosphere, presented in  \cite{smitha-2013ApJ...768..163S} and in this paper, 
helps only for $\mu=0.1$. For other $\mu$ positions, the temperature structure has to be further modified. 
Thus we are unable to find a single 1D model atmosphere that fits the CLV observations of the 
Ba~{\sc ii} D$_2$ line.\footnote{In fact the complexity of the problem is such that no single 1D model atmosphere
fits the observations of different lines, apart from observations of the same line at different 
$\mu$ positions. For example, when modeling the Cr~{\sc i} triplet
in \cite{smitha-2012A&A...541A..24S}, we had to modify the temperature structure of the FALF 
model atmosphere in the deeper layers to obtain a fit to the $(I,Q/I)$ observations at $\mu=0.15$.}

\section{Conclusions}
\label{smitha-conclusions}
In this paper we briefly describe the results of our previous paper on 
modeling the polarization profiles of the Ba~{\sc ii} D$_2$ line by taking full 
account of PRD, radiative transfer, and HFS effects. Also, we extend the same 
to model the observed CLV of the $Q/I$ profiles. 
To model the CLV profiles, we use the same approach that was used in \cite{smitha-2013ApJ...768..163S}.
We find that some of the standard model atmospheres
like the  FALF, FALC, FALA, and FALX fail to reproduce
the central peak, and instead give a central dip which is mainly due to 
PRD effects. Also, this central dip is quite sensitive to the 
temperature structure of the model atmosphere. By slightly modifying the temperature
of the standard FALX model atmosphere, we obtain a fit to both $I$ and $Q/I$ profiles. 
This new model is called the $\overline{\rm FALX}$. 
However this new model successfully fits the observations only for 
$\mu=0.1$ and fails at other limb distances. 
Hence within the sample of the model atmospheres tested by us, 
no single one dimensional, single component model atmosphere succeeds in 
reproducing the observed $Q/I$ profiles at all limb distances. This 
possibly indicates that multi-dimensional transfer effects are required to 
model the CLV observations of this line.

Finally, we show that since the Ba~{\sc ii} D$_2$ is a strong resonance line, it has damping wings that are governed by PRD. 
It is therefore necessary to take full account of PRD in order to correctly model the $Q/I$ profile. 
In contrast, the intensity profiles computed with PRD or CRD do not differ substantially from each other
through out the entire line profile.

\acknowledgments
We would like to thank Dr.~Michele Bianda for useful discussions. We acknowledge the use of HYDRA 
cluster facility at the Indian Institute of Astrophysics for 
computing the results presented in the paper. We would like to thank the Referee for a careful reading of the manuscript 
and for giving useful comments and suggestions.

\end{document}